\documentclass[sigconf]{acmart}

\AtBeginDocument{%
  \providecommand\BibTeX{{%
    \normalfont B\kern-0.5em{\scshape i\kern-0.25em b}\kern-0.8em\TeX}}}

\copyrightyear{2022}
\acmYear{2022}
\setcopyright{acmcopyright}\acmConference[WWW '22]{Proceedings of the ACM Web Conference 2022}{April 25--29, 2022}{Virtual Event, Lyon, France}
\acmBooktitle{Proceedings of the ACM Web Conference 2022 (WWW '22), April 25--29, 2022, Virtual Event, Lyon, France}
\acmPrice{15.00}
\acmDOI{10.1145/3485447.3512014}
\acmISBN{978-1-4503-9096-5/22/04}

\usepackage{graphicx} 
\graphicspath{{figures/}} 
\usepackage{multirow}
\usepackage{dsfont}
\usepackage{colortbl}
\usepackage[ruled,linesnumbered]{algorithm2e}

\usepackage{float}

\usepackage{caption}
\usepackage{subcaption}

\usepackage{balance}

\usepackage{todonotes}
\makeatletter
\newcommand*\iftodonotes{\if@todonotes@disabled\expandafter\@secondoftwo\else\expandafter\@firstoftwo\fi}  
\makeatother

\usepackage{cleveref}
\crefname{section}{\S}{\S\S}
\crefname{table}{Table}{}
\crefname{figure}{Figure}{}
\crefname{algorithm}{Algorithm}{}
\crefname{equation}{Eq.}{}
\crefname{appendix}{App.}{}
\crefname{prop}{Proposition}{}
\crefname{thm}{Theorem}{}
\crefformat{section}{\S#2#1#3}  

\newcommand{\scores}{\mathbf{r}}

\newcommand{\activetracks}{\mathbf{a}}
\newcommand{\maxactive}{\mathrm{max}(\scores_{\activetracks})}

\newcommand{\mathR}{\mathbb{R}}

\newcommand{\X}{\mathbf{X}}
\newcommand{\Y}{\mathbf{Y}}
\newcommand{\Z}{\mathbf{Z}}
\newcommand{\I}{\mathbf{I}}

\newcommand{\xxi}{\mathbf{x}_i}
\newcommand{\WQ}{\mathbf{W}^Q}
\newcommand{\WK}{\mathbf{W}^K}
\newcommand{\WV}{\mathbf{W}^V}
\newcommand{\Q}{\mathbf{Q}}
\newcommand{\K}{\mathbf{K}}
\newcommand{\V}{\mathbf{V}}
\newcommand{\OO}{\mathbf{O}}

\newcommand{\M}{\mathbf{M}}

\newcommand{\WO}{\mathbf{W}^O}

\newcommand{\Wone}{\mathbf{W}_1}
\newcommand{\Wtwo}{\mathbf{W}_2}

\newcommand{\relu}{\mathrm{ReLU}}

\newcommand{\ffb}{\mathrm{FFB}}
\newcommand{\mha}{\mathrm{MHA}}
\newcommand{\mab}{\mathrm{MAB}}

\newcommand{\isab}{\mathrm{ISAB}}
\newcommand{\LN}{\mathrm{LN}}
\newcommand{\att}{\mathrm{Att}}

\newcommand{\concat}{\mathbin\Vert}

\newcommand{\softmax}{\mathrm{softmax}}
\newcommand{\setrank}{\mathrm{SetRank}}
\newcommand{\trm}{\mathrm{SetTRM}}
\newcommand{\enc}{\mathrm{Encoder}}
\newcommand{\dec}{\mathrm{Decoder}}

\newcommand{\company}{Spotify} 

\newcommand{\sat}{\textsc{SAT}\xspace}
\newcommand{\promo}{\textsc{Boost}\xspace}
\newcommand{\ismid}{\textsc{Exposure}\xspace}
\newcommand{\disco}{\textsc{Discovery}\xspace}

\newcommand{\boost}{\textsc{Boost}\xspace}
\newcommand{\emerging}{\textsc{Exposure}\xspace}
\newcommand{\discovery}{\textsc{Discovery}\xspace}

\newcommand{\ndcgfive}{\textsc{NDCG@5}\xspace}
\newcommand{\ndcgten}{\textsc{NDCG@10}\xspace}

\newcommand{\mostra}{\textsc{Mostra}\xspace}
\newcommand{\fc}{DNN}

\newcommand{\moltr}{MO-LTR}

\newcommand{\mostrasum}{\textsc{Mostra-WtSum}\xspace}

\newcommand{\mocff}{JMO-BS}
\newcommand{\attnrank}{AttRank}

\begin{document}
\fancyhead{}

\title{\mostra: A Flexible Balancing Framework to Trade-off User, Artist and Platform Objectives for Music Sequencing}





\author{Emanuele Bugliarello}
\authornote{Part of the work conducted when Emanuele Bugliarello interned at Spotify Research.\\$^{\dagger}$Rishabh Mehrotra is now at ShareChat.}
\affiliation{
    \institution{University of Copenhagen}
    \country{Denmark}
}
\email{emanuele@di.ku.dk}

\author{Rishabh Mehrotra$^{\dagger}$ \hspace{0.01cm} James Kirk \hspace{0.01cm} Mounia Lalmas}
\affiliation{
    \institution{Spotify}
    \country{UK}
}
\email{{rishabhm,jkirk,mounial}@spotify.com}

\renewcommand{\shortauthors}{Bugliarello, Mehrotra, Kirk \& Lalmas}

\begin{abstract}
We consider the task of sequencing tracks on music streaming platforms where the goal is to maximise not only user satisfaction, but also artist- and platform-centric objectives, needed to ensure long-term health and sustainability of the platform. Grounding the work across four objectives: Sat, Discovery, Exposure and Boost, we highlight the need and the potential to trade-off performance across these objectives, and propose Mostra, a Set Transformer-based encoder-decoder architecture equipped with submodular multi-objective beam search decoding. The proposed model affords system designers the power to balance multiple goals, and dynamically control the impact on one objective to satisfy other objectives. Through extensive experiments on data from a large-scale music streaming platform, we present insights on the trade-offs that exist across different objectives, and demonstrate that the proposed framework leads to a superior, just-in-time balancing across the various metrics of interest.

\end{abstract}




\maketitle

\section{Introduction}

\begin{figure}
    \centering
    \includegraphics[width=0.95\linewidth, trim={2cm 2cm 2cm 0cm}, clip]{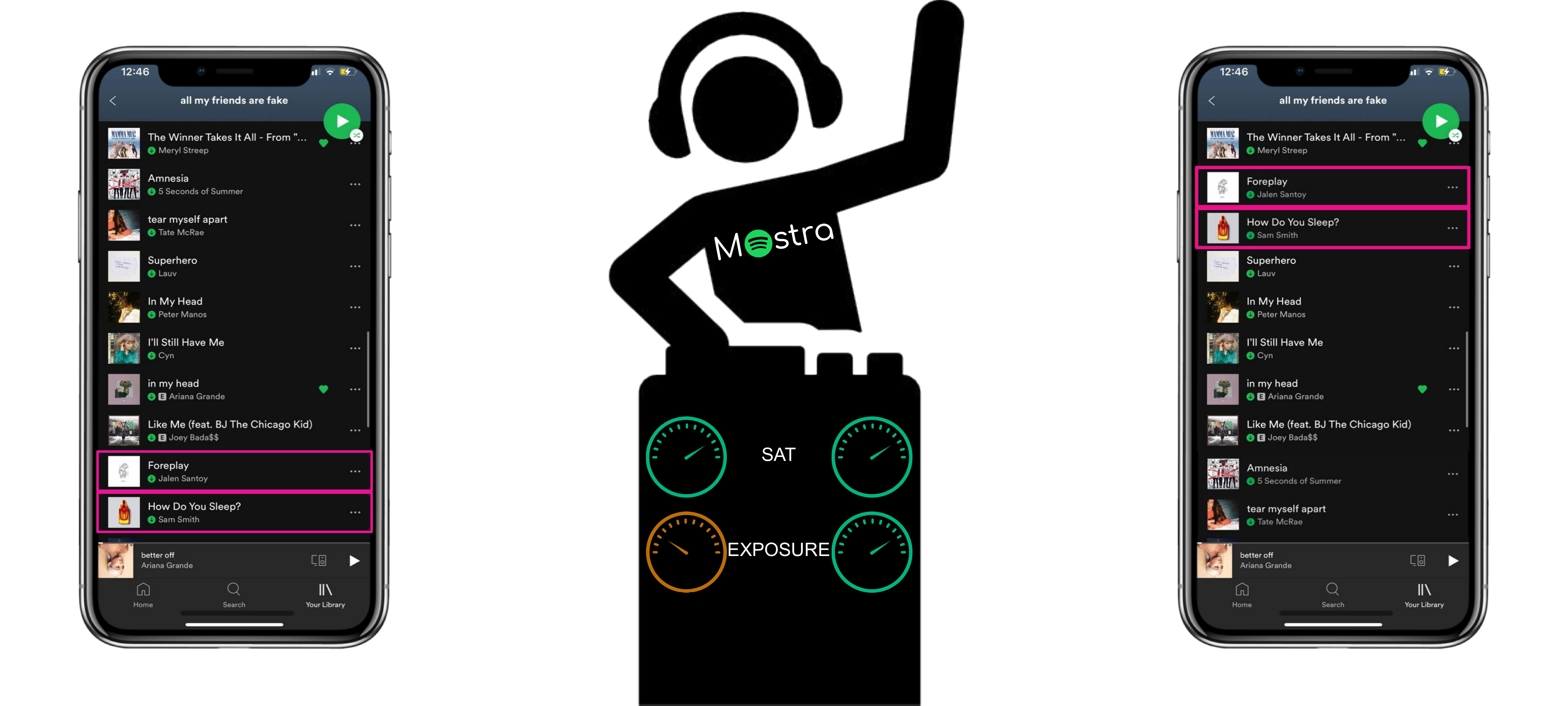}
    \vspace{-3mm}
    \caption{Music sequencing to cater to different objectives.}
    \label{fig:figure1}
    \vspace{-3mm}
\end{figure}

Recommender systems shape the bulk of consumption on digital platforms, and are increasingly expected to not only comprehend and fulfil user's immediate short-term needs, but also to help them discover new content and expand their tastes for continued long-term engagement~\cite{10.1145/3159652.3159700}. 
Beyond users, recommenders also benefit content creators and suppliers by helping them get exposed to consumers and grow their audience~\cite{Abdollahpouri2020MultistakeholderRS}. Indeed, most modern digital platforms are multi-stakeholder platforms (e.g. AirBnb: guests and hosts, Youtube: consumers and producers, Uber: riders and drivers, Amazon: buyers and sellers), and rely on recommender systems to strive for a healthy balance between user needs, exposure needs of creators and suppliers, as well as platform objectives~\cite{mehrotra2018towards}, to ensure long-term health and sustainability of the platform.

Considering music streaming platforms, a fundamental requirement of a music recommender system is its ability to accommodate considerations from the users (e.g. short-term satisfaction objectives), artists (e.g. exposure of emerging artists) and platform (e.g. facilitating discovery and boosting strategic content), when surfacing music content to users. Furthermore, different recommendation surfaces care about such objectives to different extents; e.g. a playlist like \textit{Fresh Finds} would prioritise featuring emerging up-and-coming artists, to help listeners discover music from artists they might not have otherwise come across; while \textit{All Out 90s} would prioritise nostalgic music familiar to users. This motivates the need for recommendation models that allow system designers to have explicit, on-the-fly control over the trade-offs across different objectives when designing recommendation strategies for their surfaces.

In this work, we focus on the task of multi-objective sequencing, where a model is tasked with ranking music tracks from a large set of candidate tracks to satisfy user-centric, artist-centric and platform-centric objectives. We present \mostra---\textbf{M}ulti-\textbf{O}bjective \textbf{S}et \textbf{Tra}nsformer---a set-aware, encoder-decoder framework for flexible, just-in-time multi-objective recommendations.
Our encoder builds on top of Set Transformers~\cite{pmlr-v97-lee19d}, having the ability to encode an entire input set of music tracks (i.e. a set of vectors) and capture high-order interactions between tracks through self-attention. 

We present a novel \emph{submodular multi-objective} beam search decoder that allows us to trade-off gains across multiple objectives, while considering \emph{counterfactual}\footnote{NB: Here, \emph{counterfactual} (\cref{sec:algo}) differs from the one used in the causal literature~\cite{menzies2001counterfactual}.} performance on user metrics. The inference time tuning ability (i.e. \emph{just-in-time} decisioning) of the proposed beam search decoder affords the \emph{flexibility} desired by system designers to satisfy and trade-off multiple objectives to different degrees based on dynamic, strategic needs (\cref{fig:figure1}).

Looking at music consumption data from a large-scale track sequencing framework powering \company, we find evidence around differential correlational overlap across user-, artist- and platform-centric goals. Further, we uncover vast heterogeneity across sessions in terms of objective co-occurrences, with different sessions supporting objectives to different extent, hinting at differential competition amongst objectives across sessions. Subsequently, we find evidence that user satisfaction significantly varies based on such objective overlap, further motivating the need for efficiently balancing objectives for music sequencing.

Based on extensive experiments, we demonstrate that the proposed \mostra framework is able to deliver on the above requirements, and obtains gains across artist- and platform-centric objectives without loss in user-centric objectives compared to state-of-the-art baselines. Moreover, detailed analyses indicate that our approach seamlessly adapts to sessions with different characteristics and multi-objective trade-offs. Our code is available online.\footnote{\url{https://github.com/spotify-research/mostra}.}

Taken together, our work sheds light on the tension across different stakeholder objectives in music recommendations, and equips system designers with a practical framework to design flexible recommendation policies suiting evolving strategic business needs. 
\section{Related Work}

\paragraph{Recommender systems.}

Modern day recommender systems have evolved from using matrix and tensor factorisation approaches ~\cite{10.1145/1864708.1864727,Bugliarello+:IJCAI2019,5197422}, to leveraging the modelling capabilities of deep neural networks~\cite{10.1145/3285029}.
Within this family of architectures, two main categories can be distinguished: \emph{instance-level} and \emph{set-level} models.
The former group predicts a score for each item independently of any other item recommended to the user~\cite{10.1145/3038912.3052569}.
More recently, motivated by studies showing that a user's decision about a given item can be affected by other items presented to them~\cite{10.1145/2009916.2010057}, set-level models have been investigated.
This class of learning-to-rank methods compare multiple documents together by means of multivariate scoring functions. 
In particular, $\setrank$~\cite{10.1145/3397271.3401104} is a transformer based ranking model that achieves state-of-the-art performance on popular recommendation tasks~\cite{10.1145/2987380,DBLP:journals/corr/QinL13,pmlr-v14-chapelle11a}. 
Our proposed method also relies on Set Transformers~\cite{pmlr-v97-lee19d} to preserve its input set permutation invariance, but it produces \emph{dynamic rankings} of the items via a beam search algorithm that accounts for satisfying multi-stakeholder objectives.

\vspace{-1mm}
\paragraph{Multi-objective recommendations.}
Multi-objective optimisation is a well-studied area in operation research and machine learning~\cite{NIPS2014_5588}, with recent approaches often aggregating multiple objectives into a single function~\cite{Kim2006,mehrotra2018towards}, or designing reward functions with multiple objectives~\cite{hansen2021shifting}.
Several search and recommendation applications need to meet multi-objective requirements, from click shaping~\cite{agarwal2012personalized} to email volume optimisation~\cite{gupta2016email}.
Recent approaches developed multi-objective techniques to trade-off various aspects of recommendations~\cite{10.1145/3394486.3403374, 10.1145/3442381.3449889}, provide novel and diverse recommendations~\cite{ribeiro2012pareto}, or to balance familiarity with discovery while serving recommendations~\cite{mehrotra2021algorithmic}.
Our proposed method also aims at providing diverse multi-objective recommendations by a \emph{submodular} scoring function within a \emph{counterfactual} beam search algorithm.

\vspace{-1mm}
\paragraph{Beam search algorithms.}
Beam search is an approximate search algorithm widely used in machine translation~\cite{edunov-etal-2018-understanding} and image captioning~\cite{AAAI1817329}.
As the space of all possible sequences is exponentially large, it is often intractable to find the sequence that maximises a specific objective.
Beam search is a limited-width breadth-first search algorithm that, at every step $n$, expands at most $k$ partial sequences of length $n+1$ that maximise a sequence objective. 
An extension, constrained beam search have recently been used to constrained words generation~\cite{anderson-etal-2017-guided,post-vilar-2018-fast}, and recommender systems~\cite{10.1145/3308558.3313568}. 
\mostra embeds \emph{high-quality} and \emph{diversity} into the beam search algorithm, allowing us to dynamically and \emph{efficiently} adjust recommendations at serving time in a large-scale environment.

\vspace{-1mm}
\paragraph{Controllable multi-objective recommendations.}
Two methods are of particular relevance to our work.
On the one hand, Jannach et al.~\cite{10.1145/2792838.2800182} proposed a method to generate music playlists with coherent tracks. 
It is based on finding the k-NN next tracks w.r.t. the playlist history so far.
That is, this approach focuses on similarity of tracks, and, as such, is not ideal for our scenario where satisfying long-term strategic goals requires finding music tracks that are different from the ones the users often play.
On the other hand, ComiRec~\cite{cen2020controllable} models the multiple interests of users through different, predefined categories.
This is clearly a limitation in our setup, where items (songs) can change their category (objective) every day (e.g. a song by an artist being promoted) or are user-specific (e.g. \disco songs).
Moreover, this method, during inference, recommends the most similar items across all categories, which, as noted above, is a limitation in our use case. Extending~\cite{10.1145/2792838.2800182} and~\cite{cen2020controllable} is an interesting direction for future work.\\

\vspace{-1mm}
In this paper, we present a framework that is both \emph{input} and \emph{output} set-aware. The proposed framework, \mostra,  encodes a set of music tracks at once, and efficiently decodes them while ensuring diversity across multiple objectives.
Moreover, we show, for the first time, how beam search can be adapted to provide an end-to-end, scalable solution to just-in-time, multi-objective recommendations needed in multi-stakeholder platforms.
\section{Objectives \& Stakeholders} \label{sec:objectives}
Our goal is to understand and leverage the trade-off across different user-, artist- and platform-centric goals, for music recommendation. 
We focus on the task of set sequencing: \emph{given a set of items, the goal is to order them to meet multiple objectives}. 
We begin by describing the music streaming context in which we instantiate our work, and present insights on objectives interplay across sessions that underpins the scope of objective balancing when sequencing tracks. 

\begin{figure*}[t]
    \centering
    \includegraphics[width=0.22\linewidth]{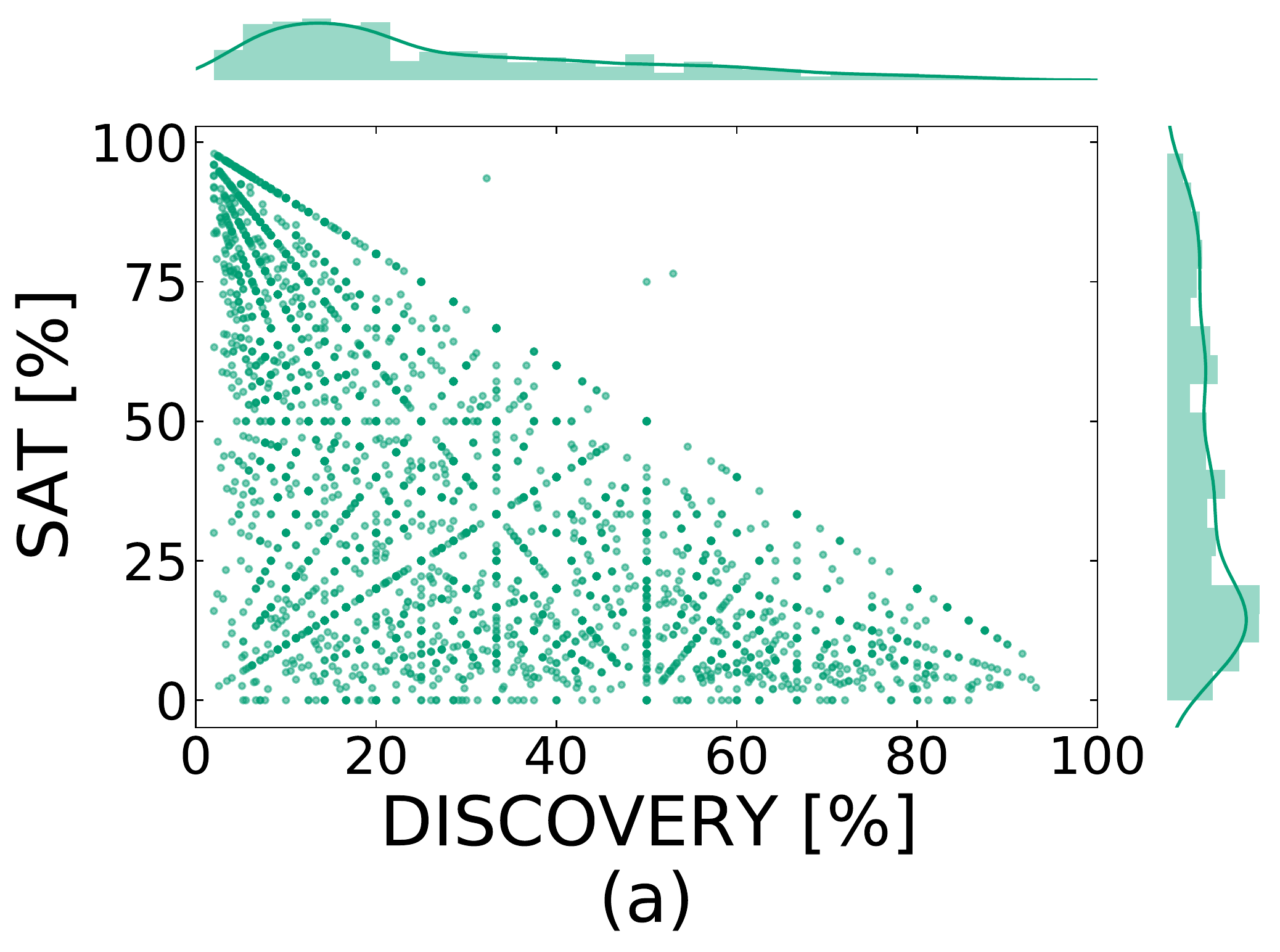}
    \includegraphics[width=0.21\linewidth]{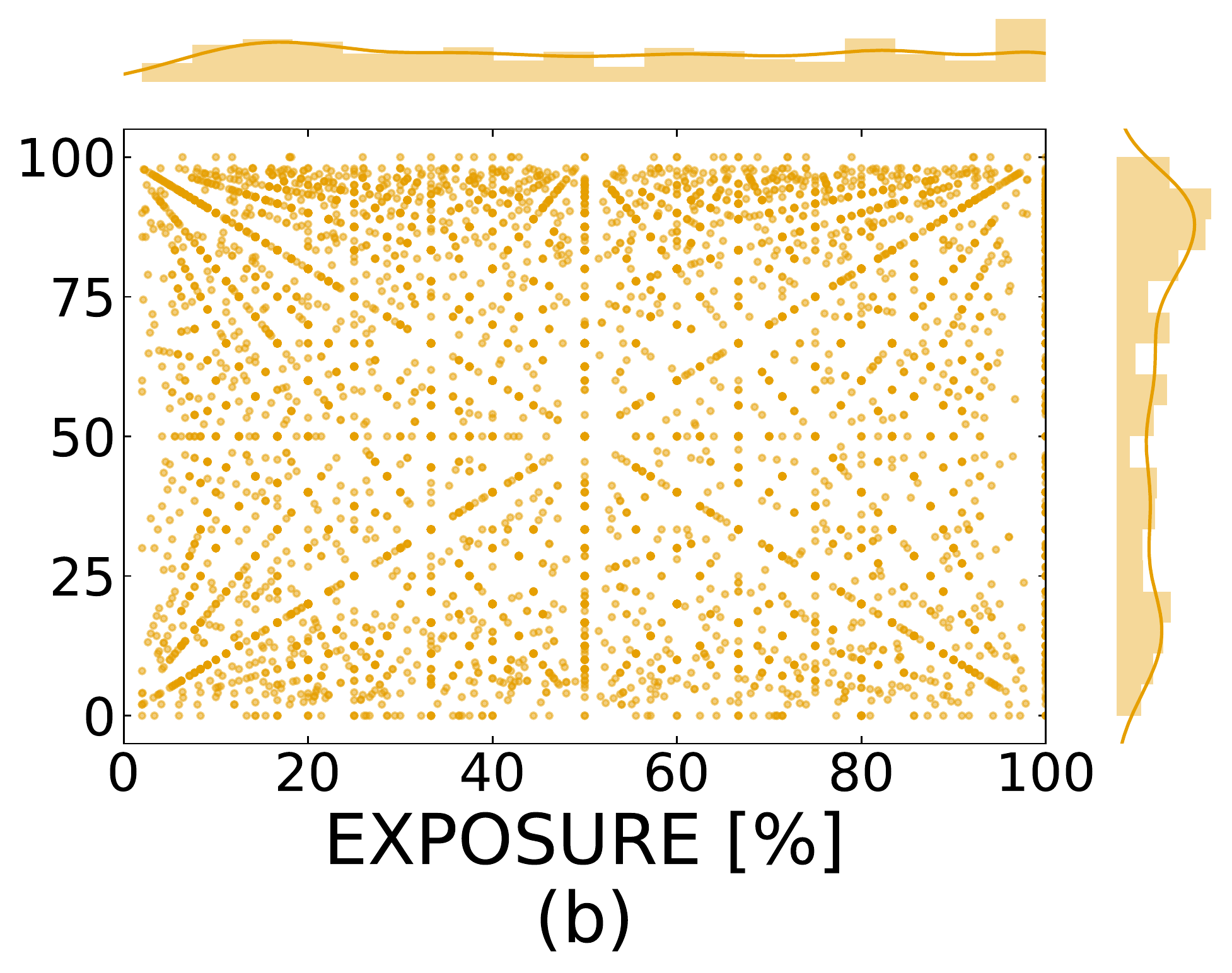}
    \includegraphics[width=0.21\linewidth]{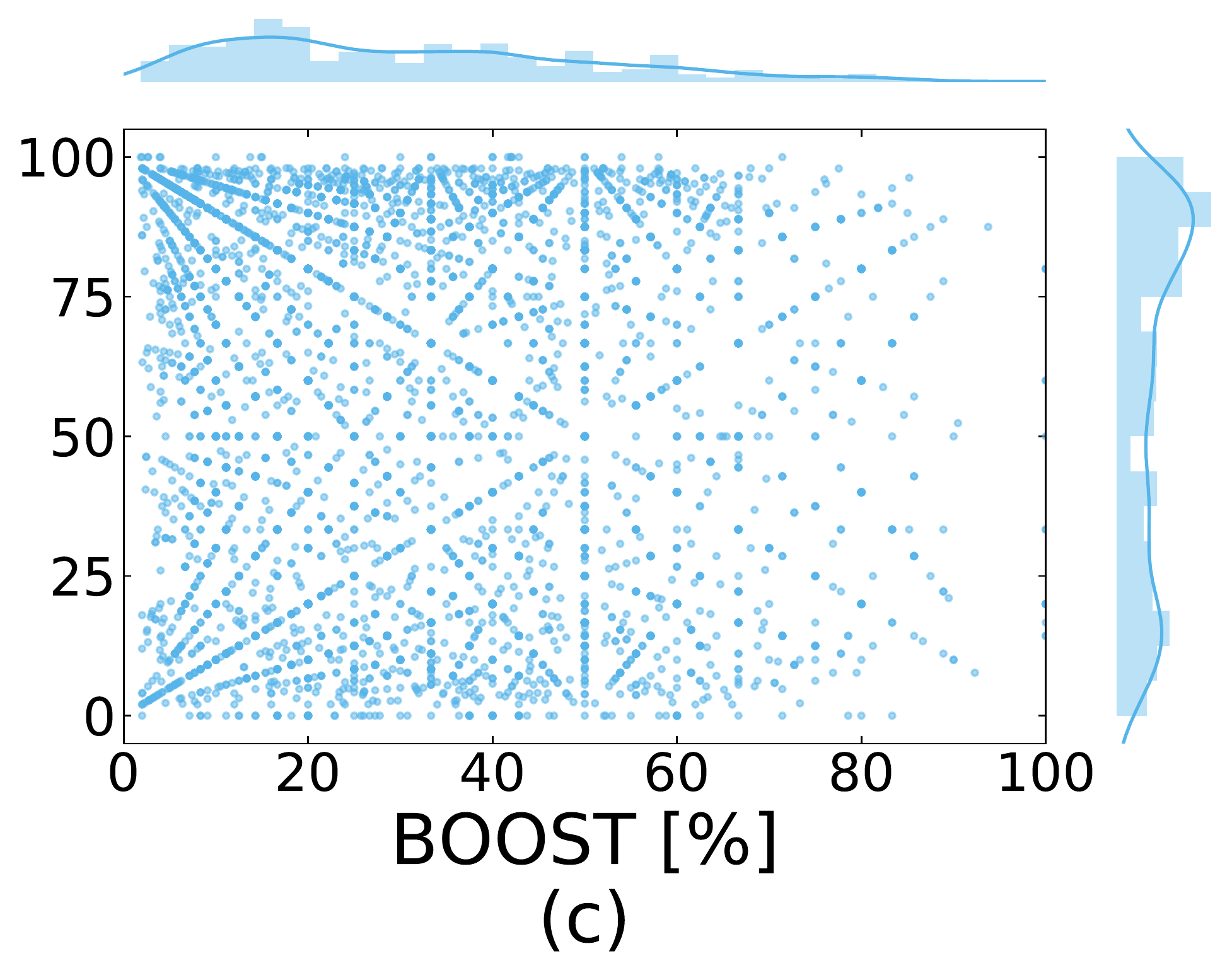}
    \includegraphics[width=0.15\linewidth]{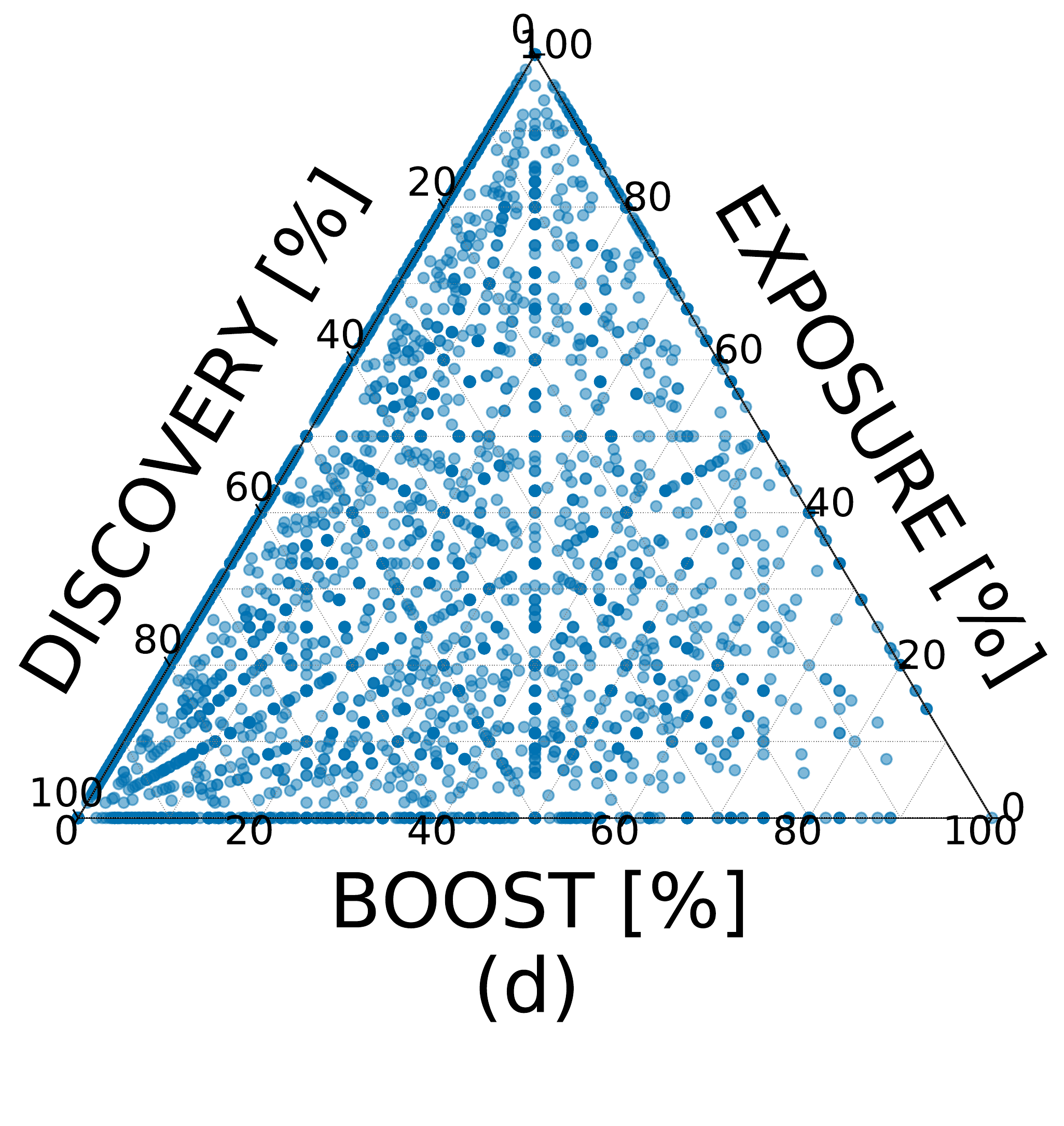}
    \includegraphics[width=0.15\linewidth, trim={0cm 0cm 0cm .2cm}, clip]{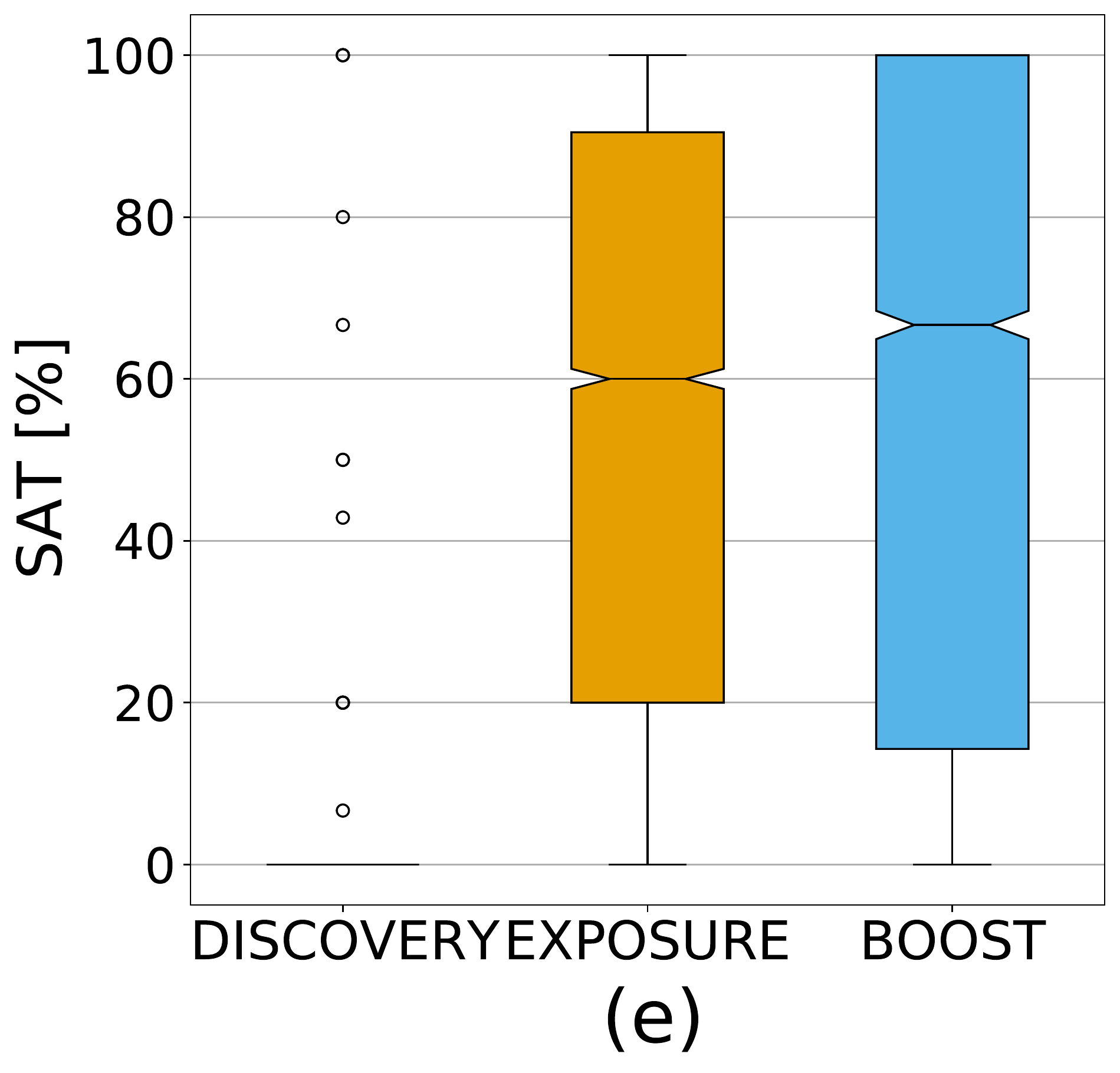}
    \vspace{-4mm}
    \caption{\small{Average set satisfaction score as a function of the percentage of tracks with platform-centric objectives (a, b, c). Ternary plot of set percentage of attribute tracks (d). Boxplot of average user satisfaction for platform-centric tracks (e).}}
    \label{fig:heatmap}
     \vspace{-2mm}
\end{figure*}

\subsection{Data Context}
We consider the specific use case of \company, a global music streaming platform wherein a recommender system is tasked with generating a playlist from a set of available tracks.  Our dataset consists of the listening history of a random subset of $10$ million distinct users with over 500M sessions resulting in over $1$ billion interactions during a $7$ day period. We focus on radio-like, music streaming sessions. 
Each track could come from different artists and have different characteristics when paired with a given user. For each user--track pair, we assume access to four binary scores that report whether the pair satisfied any of the objectives defined next.

\subsection{Objective Definitions}\label{sec:objective-defs}

Given that recommender systems shape content consumption, they are increasingly being optimised not only for user-centric objectives, but also for objectives that consider supplier needs and long-term health and sustainability of the platform. Blindly optimising for user relevance only has shown to have a detrimental impact on supplier exposure and fairness~\cite{mehrotra2018towards}. 
Also, there often exists content that is not surfaced to users, due a lack of metadata, popularity bias or information needs that are hard for users to articulate. 
Furthermore, users often rely on such systems to discover content which they may be less familiar with, thereby gradually evolving and expanding their music taste.
To this end, we consider a multi-dimensional approach to recommendations, and consider a number of different objectives:

\begin{description}
    \item [Short-term satisfaction:] Recommender systems want to keep users satisfied, and thus recommend content and services tailored to users based on knowledge about their preferences and behaviour. We consider \emph{track completion} as a short-term user satisfaction metric, a binary value reporting whether a user streamed a track or skipped it. We use it as a proxy for user satisfaction and denote it as \sat.
    
    \item [Music discovery:] Users often have access to large repositories of music content with only a small fraction familiar to them. Music \textit{discovery} is the process that allows users to experience content previously unknown to them, and has been shown as as main motivation to continue platform subscription~\cite{mantymaki2015gratifications} and an important need for music listeners~\cite{garcia2018understanding,lee2016users}. In this work, a track is labelled as a \discovery for a given user if that user has never listened to that track or any tracks produced by the track's artist. Platforms would want to anticipate and serve user's discovery needs, and as such, we consider facilitating \discovery as an explicit platform-centric objective.
    
    \item [Emerging artist exposure:] Helping emerging artists build an audience allows the platform to support a wide variety of creators, and is an important consideration for sustainability of two-sided platforms. 
    We explicitly consider \emph{exposure} of emerging artists as an artist-centric objective, which a recommender system may consider when making recommendation decisions. 
    An artist is emergent if their popularity, for example, lies in the inter-quartile range.
    A given track is labelled as \ismid if the track's artist is considered as an \emph{emerging} artist by the platform.
    
    \item [Content boosting:] There might be several other reasons why a platform might wish to boost consumption of specific content at a give time, e.g. based on recent world events, to align with certain trends, or to celebrate and honour certain artists. We consider a content \emph{boosting} objective to denote a platform-centric objective wherein the platform intends to boost consumption of content of strategic importance. 
    In our setup, each track is assigned a binary \boost value reporting whether it can be boosted or not.
\end{description}

\noindent
These objectives are available to the recommender system; they are linked to each user--track pair by extracting them from the historic interaction data (e.g.~\discovery) or through editorial annotations (e.g.~\boost).
Finally, while these objectives are specific to music streaming platforms, our objective definitions, our setup and  framework are amenable to other interpretations and definitions, and applicable in a wider variety of industrial settings~\cite{mehrotra2019recommendations}.

\subsection{Understanding Objectives within Sets}
To better understand the interplay between the four objectives in user listening sessions, and how frequent some of these objectives are within sessions,\footnote{\noindent\textbf{Note on sets and sessions:} we interchangeably use these two terms.} we consider a random sample of $100$M listening sessions and present insights on the observed relationships. 

In~\cref{fig:heatmap}(a,b,c), we compute the average user satisfaction (i.e. average of track completion rate across all tracks) and plot this against the percentage of tracks in that session belonging to the three other objectives, \discovery, \emerging and \boost, respectively. 
Both \emph{exposure} to emerging artists and \emph{boosting} objectives are not correlated to our user-centric objective, \sat, while our \emph{discovery} objective is negatively correlated with it: the higher the percentage of discovery tracks in a set, the lower the user satisfaction. 
This highlights that optimising for discoveries might have a \emph{detrimental} effect on our user-centric metric, and the recommender system will need to find a \emph{trade-off} between these two objectives. 

To investigate how often these objectives co-occur in user sessions (and correspondingly in candidate sets), we plot the distribution of artist- and platform-centric objectives across sampled sets in~\cref{fig:heatmap}(d). The diagram clearly demonstrates the vast diversity of set types in our data: some sessions only have tracks belonging to one of these objectives, while a significant number of sets have tracks belonging to each of these objectives. 
Moreover, looking at the distribution of the objectives (histograms at the top of scatter-plots in \cref{fig:heatmap}(a,b,c)), we see that the percentage of tracks belonging to emerging artists (\emerging) is uniformly distributed, while most of the sets only have a small portion of \boost and \discovery tracks. 
Thus, we anticipate more severe competition across objectives in certain sessions than others, due to competing objectives. 
This also motivates the need to incorporate \emph{set-awareness} into the recommender model, so that it can adaptively find better solutions based on the objective composition of specific input sets.

Finally, in~\cref{fig:heatmap}(e), we show the distribution (as a box plot) of user satisfaction of each artist- and platform-centric objective. 
Here, we see that the user satisfaction may vary significantly depending on which specific \promo and \ismid tracks are shown.
On the other hand, \disco tracks usually lead to user dissatisfaction in our sample data. 
Nevertheless, there are instances in which even these tracks lead to high \sat scores. 
This plot clearly shows that cleverly selecting artist- and platform-centric tracks is key to obtain relevant recommendations for the users.

These analyses motivate us to develop multi-objective recommendation models that not only take into account the objective composition of different candidate sets, but also consider the interplay between user satisfaction and other stakeholder objectives.
\section{End-to-End Neural Architecture} \label{sec:model}

We introduce our approach for just-in-time multi-objective recommendations, called \mostra (\textbf{M}ulti-\textbf{O}bjective \textbf{S}et \textbf{Tra}nsformer). We begin by defining the mathematical notations.

\subsection{Notation}\label{sec:model-a}

Plain letters denote scalars ($a$); bold lower-case letters denote vectors ($\mathbf{a}$) and bold upper-case letters denote matrices ($\mathbf{A}$). 
$\mathbf{a}_i$ represents the $i$-th element of $\mathbf{a}$, while $\mathbf{A}_i$ is the $i$-th row of $\mathbf{A}$.
We consider a unified view of neural recommender systems as encoder-decoder architectures~\cite{cho-etal-2014-learning}:
\begin{equation}
    \mathrm{RecSys}(\X) = \dec \circ \enc(\X),
\end{equation}
\noindent where (i) the encoder maps an input set (e.g.~tracks) into an embedding space, and (ii) the decoder returns scalars (e.g.~probability of track completion) that can be used to generate an ordered sequence.
Overall, given an input set of $N$ $d$-dimensional tracks $\X\in\mathR^{N\times d}$, a neural recommender system (RecSys) is a composition of an encoder and a decoder functions.
%

\begin{figure*}
    \centering
    \includegraphics[width=0.95\linewidth, trim={0 11cm 8cm 0}, clip]{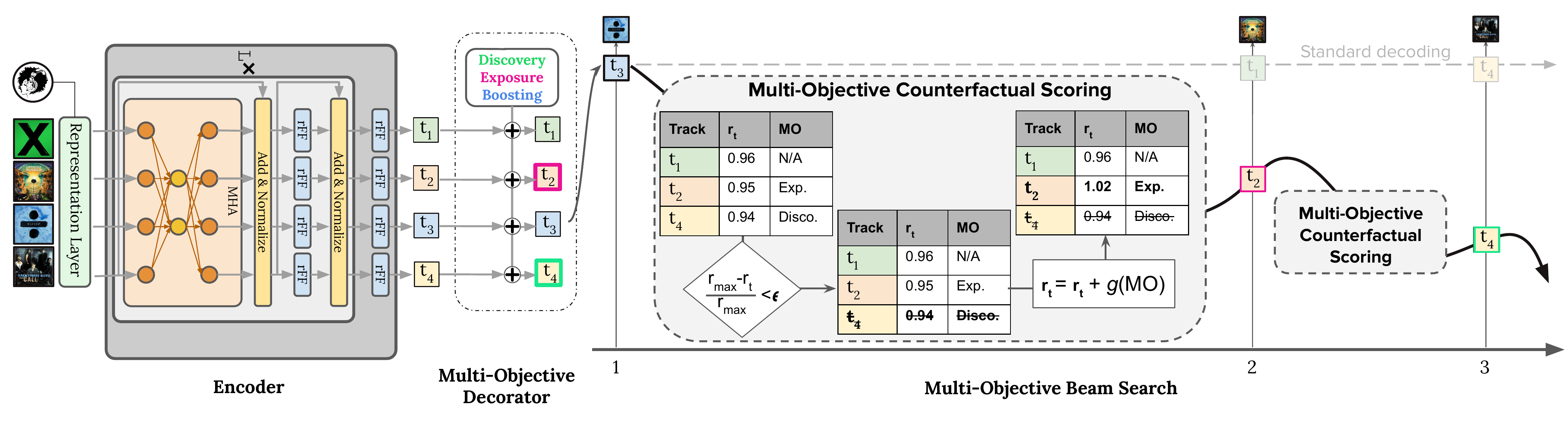}
    \vspace{-6mm}
    \caption{\small{Diagram of the overall architecture: user and track features are extracted and fed into a Set Transformer-based encoder that predicts relevance scores. At inference, additional artist- and platform-centric objectives are considered by a multi-objective counterfactual beam search algorithm to sequence tracks, simultaneously maximising all goals.}}
    \label{fig:model}
    \vspace{-2mm}
\end{figure*}

\subsection{Overview of \mostra Architecture} \label{sec:model-b}

\cref{fig:model} shows the overall proposed end-to-end neural architecture for multi-objective track sequencing, consisting of three main parts.

First, a \textbf{representation layer} encodes derived user embeddings, learnt track embeddings, and joint user--track features to generate input representations. 
Then a \textbf{Set Transformer-based encoder} considers the whole set of tracks (i.e., a set of vectors) as input, and employs a stack of multi-head self-attention blocks to model high-order interactions between tracks to jointly learn embeddings for the whole set.

The encoded contextualised representations are then mapped into relevance scores (scalars) by means of a feed-forward layer. 
These scores denote the predicted user satisfaction for each user--track pair. 
Finally, a \textbf{multi-objective beam search decoder} leverages the predicted encoder scores, with additional artist-centric and platform-centric decorated inputs for each user--track pair.
It employs a flexible, submodular scoring technique to produce a dynamic track recommendation sequence that balances user satisfaction and multi-objective requirements at a given time.

\subsection{Representation Layer} \label{sec:model-c}

Each track is represented as a concatenation of three distinct feature vectors: a \emph{contextual} vector, an \emph{acoustic} vector, and a \emph{statistic} vector. 
The {contextual} vector is a multi-dimensional real-valued vector, trained such that two tracks that occur in the same context are close in the vector space~\cite{mehrotra2018towards}. 
The {acoustic} vector consists of $16$ derived features that reflect acoustic properties of the track, e.g., loudness. Last, the {statistics} vector contains information on the track length and its popularity. 
See \cref{sec:exp-details} for an overview of the features.

Each user is represented as a weighted average of the {contextual} vectors of the tracks the user has played in the past, as described in~\cite{mehrotra2018towards}. 
For each user and track pair, there are a number of derived features capturing their relations, such as the cosine similarity between the user's and track's {contextual} vectors. 
Additionally, each user has an affinity for all genres, which is used as a feature by taking the maximum affinity within the track's genres. 

Finally, input tracks $\mathcal{T} = \{\mathbf{t}_1, \mathbf{t}_2, \dots, \mathbf{t}_N\}$ and corresponding user vector $\mathbf{u}$ are combined to produce $d$-dimensional input embeddings:
\begin{equation} \label{eq:feats}
    \xxi = \phi(\mathbf{u}, \mathbf{t}_i),
\end{equation}
where $\phi$ denotes the function for feature extraction.

\subsection{Set Transformer Encoder} \label{sec:model-d}

One of the core characteristics of our proposed \mostra architecture is its ability to consider the entire set of tracks and develop a multivariate scoring function that scores each permutation of the original set of tracks. 
Unlike traditional, item-level neural architectures, multivariate scoring allows us to model and compare multiple tracks together, capture the local context information and account for cross-track interaction effects. 
To this end, we base our encoder on the Set Transformer model~\cite{pmlr-v97-lee19d}, a variant of the popular Transformer architecture~\cite{vaswani2017attention} to handle set-input problems. We now detail the components of our Set Transformer-based encoder.

\paragraph{Transformer layer.}
Let $\X\in\mathR^{N\times d}$ be the input track representations. A Set Transformer layer encodes them by two sub-layers: an induced self-attention block (ISAB) and a feed-forward block (FFB).

\paragraph{Induced self-attention block.}
An induced self-attention block is an extension of the standard multi-head attention block (MAB) used in standard Transformer layers and defined as:
\begin{equation} \label{eq:mab}
    \mab(\Y, \Z) = \LN(\Y + \mha(\Y, \Z, \Z)),
\end{equation}
where $\LN$ is layer normalisation~\cite{ba2016layer}, and $\Y$ and $\Z$ denote internal representations of the inputs $\X$.
$\mha$ denotes a multi-head attention function that projects internal representations -- commonly known as query ($\Q$), key ($\K$) and value ($\V$) matrices -- into $H$ different matrices before computing the attention ($\att$) of each projection, concatenating ($[\cdot\concat\cdot]$) them together and mapping them with a linear transformation $\WO$.
Formally, $\mha$ is defined as:
\begin{equation} \label{eq:mha}
\begin{array}{c}
    \mha(\Q, \K, \V) = [\OO_1 \mathbin\Vert \cdots \mathbin\Vert \OO_H]\WO, \\
    \text{where } \OO_h = \att\left(\Q\WQ_h, \K\WK_h, \Q\WV_h\right),\\
    \text{and } \att(\Q_h, \K_h, \V_h) = \omega\left(\Q_h\K_h^\top\right)\V_h.
\end{array}
\end{equation}
Here, $\omega$ denotes a row-wise, scaled softmax: $\omega_i(\cdot) = \softmax(\cdot/\sqrt{d})$, and $\{\WQ_h, \WK_h, \WV_h \}^H_{h=1}$ and $\WO$ are learned matrices. 
An induced self-attention block with $M<N$ \emph{trainable} vectors $\I\in\mathR^{M\times d}$ ($\isab_M$) is then defined as:
\begin{equation} \label{eq:isab}
\begin{array}{c}
    \isab_M(\X) = \mab(\X, \mathbf{H}), \\
    \text{where } \mathbf{H} = \mab(\I, \X).
\end{array}
\end{equation}
The vectors $\I$ are called inducing points and act as low-rank projections of $\X$, allowing the model to efficiently process input sets of any size.
Moreover, an $\isab$ sub-layer is permutation equivariant~\cite{pmlr-v97-lee19d}.

\paragraph{Feed-forward block}
In each layer, the $\isab$ output representations ($\M\in\mathR^{N\times d}$) are fed into a feed-forward block given by: 
\begin{equation} \label{eq:ffb}
    \ffb(\M) = \LN(\M + \relu(\M\Wone)\Wtwo),
\end{equation}
where $\Wone,\Wtwo^\top\in\mathR^{d\times d_{\mathit{ff}}}$ are learnable matrices.

\paragraph{Stack.}
Since each Set Transformer layer is permutation equivariant, a stack of $L$ Set Transformer layers is permutation equivariant as well.
We refer to a stack of Set Transformer layers as $\trm$.

\paragraph{Mapping layer.}
The final layer of our Set Transformer-based encoder is a row-wise feed-forward network $\mathrm{rFF}: \mathR^d\rightarrow\mathR$ that projects the $d$-dimensional output representations of $\trm$ into real values corresponding to relevance scores. 
Hence, we have:
\begin{equation}\label{eq:enc}
    \mostra_{enc}(\X) = \mathrm{rFF} \circ \trm(\X).
\end{equation}

Given user--track representations, our Set Transformer-based encoder (\cref{eq:enc}) extracts cross-track user satisfaction scores for each input track.
These scores are then passed to a beam search decoder, which combines them with additional artist- and platform-centric decorations to provide multi-objective recommendations.

\subsection{Decoding for Multi-Objective Trade-offs} \label{sec:model-e}

Our goal is to recommend a sequence of tracks that balances user satisfaction with artist and platform objectives.
As business strategic needs might warrant us to deliver different objectives at different times, or across different recommendation surfaces, we want system designers to have control over the extent of trade-off across the \emph{dynamic} objective needs during inference.

We present a decoder module that allows us to not only satisfy this requirement, but also to fulfil key constraints in large-scale recommendation systems.
Our decoder is based on beam search, a parameter-free dynamic algorithm for tight real-time constraints, which we extend to satisfy on-the-fly, multi-stakeholder needs.
We propose to balance multiple goals via a multi-objective scoring function that limits deviance from pure user-centric sequences through counterfactual decisioning: instead of selecting the track with the highest predicted relevance estimate, the model considers all tracks \emph{within} a specified bound of highest estimates, and uses artist and platform objectives to decide which track to recommend.
Note that our definition of counterfactual decision making (\cref{sec:algo}) deviates from the more common counterfactual causation~\cite{menzies2001counterfactual} as it defines a worst-case-aware approach: how much of a drop in satisfaction would we expect if we recommend a track that satisfies multiple objectives, versus one purely based on best predicted satisfaction.

Compared to standard beam search, our just-in-time multi-obje\-ctive beam search (\mocff) takes artist and platform objective scores ($\mathbf{E}$) as inputs, and relies on two functions (a counterfactual and a multi-objective one) to generate beam sequences that best balance multiple objectives.
At each decoding step $n$, the algorithm first \emph{limits} its search via a counterfactual step.
Here, tracks with low predicted satisfaction scores are discarded to prevent the subsequent multi-objective scoring function to recommend tracks with low user satisfaction.
Second, the score of each track is combined with the score of its history (beam), similar to standard beam search for sequence-to-sequence problems.
Finally, a multi-objective score is computed for each track based on its decorations to \emph{balance} its current beam score for multiple objectives.
By doing so, the algorithm re-scores tracks satisfying artist- and platform-objectives while, thanks to the counterfactual step, ensuring comparable user satisfaction as the best single-objective track available at step $n$.

We now present the details of our algorithm.
In the following, let $\scores\in\mathR^N$, $\scores_i\in[0, 1]$, represent the vector of relevance scores predicted by the encoder (\cref{eq:enc}), and let $\mathbf{E}$, $\mathbf{E}_{ij}\in\{0, 1\}$, be the $N\times J$ binary matrix of $J$ artist- and platform-centric objectives.
Moreover, for a given beam and at a decoding step, we use $\mathbf{y}$ to denote the sequence of previously recommended tracks, while $\activetracks$ denotes the available ones (i.e. tracks in the set not yet recommended). 

\subsubsection{Counterfactual Scoring} \label{sec:algo}

Counterfactual reasoning is the property of a system to anticipate changes in future events given a counterfactual decision. That is, a system reasons about possible outcomes when doing action $B$ instead of action $A$.
Our counterfactual scoring is inspired by this task, and aims at making counterfactual decisions by selecting a track based on additional artist or platform objectives, rather than purely based on user-centric objective.
Specifically, we take a worst-case-aware approach that lets the model \emph{control} the degree of sub-optimal recommendations at any decoding step, limiting the potential loss in user satisfaction.

Formally, given relevance scores $\scores$ output by our encoder (\cref{eq:enc}), when selecting a track at position $n$, only a subset $\mathbf{a}$ of $\textit{N-n}$ tracks is \emph{available} for recommendation.
Let $\mathbf{r}_\mathbf{a}$ denote their corresponding relevance scores and let $\hat{\mathbf{r}} = \maxactive$ be the highest predicted relevance score by the encoder.
Then, for any track $t\in\activetracks$, we compute its relative difference to the best next track's score as: $\delta_t = \frac{\hat{\mathbf{r}} - \scores_t}{\hat{\mathbf{r}}}, \forall t\in\activetracks.$
These scores estimate the counterfactual, i.e. what would be the relative drop in model performance for track $t$ had the model recommended a track solely based on user satisfaction objective.
We then use this estimate to control for the worst-case performance by selecting a threshold $\epsilon$ for allowed relative drop in satisfaction score.
A track $t$'s score is set to $-\infty$ if it drops performance by more than $\epsilon$:
\begin{equation}\label{eq:eps}
    \mathbf{s}_t = \begin{cases} \scores_t, & \mbox{if }\delta_t < \epsilon \\ -\infty, & \mbox{otherwise} \end{cases}
\end{equation}
This counterfactual step allows us to only consider tracks that are at most $\epsilon$ worse than the next best track, hence introducing a worst-case-aware scenario for making multi-objective decisions.

\subsubsection{Submodular MO Beam Scoring}

To encourage the model to select tracks from across different objectives, we rely on submodular diversification across objectives. 
A submodular set function is a function with diminishing returns when adding new items to a set. 
In our case, it scores tracks higher if they meet objectives not yet well represented in the beam generated so far. 
Specifically, for any available track $t$, we compute a multi-objective score measuring the diversity of objectives as:
\begin{equation} \label{eq:submod}
    \mathbf{s}_t = \mathbf{s}_t + \frac{1}{n} \sum_{j=1}^J \sqrt{\sum_{y \in [\mathbf{y}\concat t]} \mathbf{E}_{yj}}, \forall t\in \activetracks.
\end{equation}

Indeed, selecting three tracks from two objectives is preferred over selecting all from one objective: $\sqrt{2}+\sqrt{1} > \sqrt{3}+\sqrt{0}$.
Together with the counterfactual step, this lets us balance the contributions of user-centric as well as artist- and platform-centric objectives while preferring tracks that satisfy multiple objectives.

\vspace{3mm}

\noindent\textbf{Robustness of inference-based sequencers to dynamic needs.}
It is important to note that inference-based methods (e.g. beam search) give us valuable dynamic control in the face of changing business strategy needs. 
Indeed, while learnt systems need to be re-trained (and their objectives re-weighted) with changes in strategic needs, inference-based methods provide finer control over track sequencing and can incorporate such changes more easily.

\section{Empirical Evaluation} \label{sec:exps}

We conduct large-scale experiments to compare model performance on balancing user satisfaction with other artist- and platform-centric goals. 
We experiment on large-scale user interaction data from Spotify, composed of logged feedback data from live traffic from a random sample of users in a 7-day period, spanning over 1B user interactions across a random sample of $500$M radio sessions, spread over 10M users. 
The users and their interactions were split into training, validation and test sets, with no user in any two splits. 
For each user's radio session, we log all tracks the user interacted with, along with user interaction signal for each track (e.g. whether it is skipped). 
\cref{sec:exp-details} lists the logged features for interacted tracks.

\subsection{Baselines}

We compare \mostra with three groups of state-of-the-art recommendation models. 
We report details about each model in \cref{sec:exp-details}.
The first group corresponds to track-level methods, which focus on predicting scores one track at a time. 
These single-objective methods solely focus on user satisfaction, but which might surface tracks from other objectives organically, without explicit optimisation. 
\begin{description}
    \item [Relevance ranker:] Ranks tracks by estimating user--track relevance scores using the cosine similarity between contextual embeddings for each user and for each track.

    \item [\fc:] A track-level fully-connected architecture to predict user--track interaction. The network was trained with either RMSE, BCE or \attnrank{} loss functions. This set of models constitutes a well-established neural baseline~\cite{45530,10.1145/3038912.3052569,10.1145/2988450.2988454}.
\end{description}
The second group are single-objective set-level models and model interaction effects across all tracks in the set and output a permutation. 
These models were trained using the \attnrank{} objective.
\begin{description}
    \item [SetTRM2GRU:] A $\trm$ encoder with a GRU decoder.
    \item [SetTRM2TRM:] A $\trm$ encoder with a TRM decoder.
    \item [SetRank:] A state-of-the-art recommender system based on $\trm$ that sorts a set by predicted relevance scores~\cite{10.1145/3397271.3401104}.
    In our framework, it can be viewed as applying a \emph{sort} function to the input tracks according to the scores given by \cref{eq:enc}.
\end{description}
Finally, the third group is made of multi-objective sequencing methods, which operate at the set-level and considers multiple objectives.
\begin{description}
    \item [\moltr:] A learning to rank model, which extends $\setrank$ to multi-objective training. Multi-objective labels are given as input to the model, which is then trained to predict the sum of \sat and the other objectives when the track is completed: $s_i = \sat_i + \sat_i\sum_{j=1}^J \mathbf{E}_{ij}$.
    This is our learnt multi-objective baseline. 

    \item [\mostrasum:] A first, non-counterfactual version of our \mostra method, in which multi-objective labels are given at inference. The model ranks tracks according to the weighted sum of all the objectives whenever a track has a predicted relevance score higher than $0.5$, hence mimicking \moltr{} but as an inference method: $s_i = \scores_i + \mathds{1}\{\scores_i > 0.5\} \alpha\sum_{j=1}^J \mathbf{E}_{ij}$.
    This is our online multi-objective baseline. 

    \item [\mostra:] Our $\trm$-based model with multi-objective counterfactual decoding (\cref{sec:model}).
\end{description}

\begin{table}[t]
    \centering
      \footnotesize
      \vspace{-3mm}
      \setlength\tabcolsep{2pt}
      \caption{Comparison across objectives. Single/double arrows denote statistical significance (paired t-test) at $p<0.05/0.01$ after Bonferroni correction.} \label{tab:results}
      \vspace{-3mm}
      \resizebox{0.48\textwidth}{!}{%
      \begin{tabular}{l||c|ccr||c|ccr} 
        \toprule
        \multicolumn{1}{c||}{\textbf{Method}} & 
        \multicolumn{4}{c||}{\textbf{\ndcgfive [\%]}} &
        \multicolumn{4}{c}{\textbf{\ndcgten [\%]}} \\
         & 
        \multicolumn{1}{c|}{\sat} & 
        \multicolumn{1}{c}{\promo} & \multicolumn{1}{c}{\sc Expos.} & \multicolumn{1}{c||}{\sc Disco.} &
        \multicolumn{1}{c|}{\sat} & 
        \multicolumn{1}{c}{\promo} & \multicolumn{1}{c}{\sc Expos.} & \multicolumn{1}{c}{\sc Disco.} \\
        \midrule
        \multicolumn{9}{c}{Track-level Single-objective} \\
        \hline
        Relevance ranker & 63.95 & 28.01 & 51.90 & 6.03 & 71.66 & 34.53 & 58.34 & 9.36 \\
        \fc$_{\mathrm{\attnrank}}$ &  69.14 & 27.28 & 50.03 & 17.29 & 76.16 & 33.84 & 56.93 & 23.08 \\
        \fc$_{\mathrm{BCE}}$ &  69.48 & 28.68 & 45.35 & 15.87 & 75.90 & 35.10 & 53.27 & 21.67   \\
        \fc$_{\mathrm{RMSE}}$ &  69.54 & 28.35 & 45.06 & 15.80 & 75.90 & 34.79 & 52.97 & 21.63 \\
        \midrule
        \multicolumn{9}{c}{Set-level Single-objective} \\
        \hline
        $\trm$2GRU$_{\mathrm{\attnrank}}$  & 68.19 & 27.68 & 51.25 & 21.51 & 74.88 & 34.23 & 57.82 & 25.90  \\
        $\trm$2TRM$_{\mathrm{\attnrank}}$  & 65.43 & 27.45 & 50.75 & 19.62 & 72.88 & 34.05 & 57.52 & 24.91  \\
        \rowcolor[gray]{.9} $\setrank$ & 68.78 & 28.95 & 45.64 & 17.04 & 75.37 & 35.28 & 53.50 & 22.61 \\
        $\setrank_{\mathrm{BCE}}$  & 65.47 & 26.67 & 52.56 & 20.59 & 72.79 & 33.52 & 59.03 & 25.55  \\
        $\setrank_{\mathrm{RMSE}}$ & 70.11 & 28.94 & 44.81 & 15.48 & 76.37 & 35.21 & 52.87 & 21.39 \\
        \midrule
        \multicolumn{9}{c}{Set-level Learnt Multi-objective} \\
        \hline
        \moltr$_{\mathrm{\attnrank}}$  & 54.63 & 27.19 & 54.12 & 48.76 & 64.70 & 33.87 & 60.26 & 48.76  \\
        \moltr$_{\mathrm{RMSE}}$  & 53.74 & 40.42 & 70.09 & 48.77 & 64.07 & 44.75 & 72.94 & 48.77  \\
        \midrule
        \multicolumn{9}{c}{Set-level Online Multi-objective} \\
        \hline
        \mostrasum{} $(\alpha=1.0)$  & 61.05 & 52.03 & 70.58 & 34.61 & 69.47 & 53.81 & 72.92 & 36.75  \\
        \mostrasum{} $(\alpha=0.5)$  & 61.19 & 51.86 & 70.37 & 34.31 & 69.58 & 53.67 & 72.76 & 36.50  \\
        \mostrasum{} $(\alpha=0.1)$  & 64.91 & 45.16 & 61.45 & 26.91 & 72.40 & 48.20 & 65.59 & 30.48  \\
        \mostrasum{} $(\alpha=0.01)$  & 69.49 & 31.43 & 47.18 & 16.92 & 75.88 & 37.18 & 54.62 & 22.50 \\
        \rowcolor[gray]{.9} \mostra $(\epsilon=0.01)$  & 69.56$^\Uparrow$ & 31.08$^\Uparrow$ & 46.56$^\Uparrow$ & 16.77 & 75.94$^\Uparrow$ & 36.90$^\Uparrow$ & 54.17$^\Uparrow$ & 22.37$^\downarrow$  \\
        \mostra $(\epsilon=0.05)$  & 67.66$^\Downarrow$ & 36.76$^\Uparrow$ & 51.14$^\Uparrow$ & 21.13$^\Uparrow$ & 74.60$^\Uparrow$ & 41.09$^\Uparrow$ & 57.39$^\Uparrow$ & 25.48$^\Uparrow$   \\
        \mostra $(\epsilon=0.10)$  & 66.33$^\Downarrow$ & 39.83$^\Uparrow$ & 53.86$^\Uparrow$ & 24.00$^\Uparrow$ & 73.54$^\Uparrow$ & 43.74$^\Uparrow$ & 59.65$^\Uparrow$ & 27.85$^\Uparrow$   \\
        \bottomrule
      \end{tabular}%
      }
    \vspace{-6mm}
\end{table}

We train each model for 12,500 steps on one T4 GPU using the Adam optimiser with a learning rate of 5e-4 and decay factor of 0.1.
We batch 128 sequences, each with up to 100 tracks.
Model selection was based on the best validation performance.
We use NDCG~\cite{10.1145/582415.582418} to evaluate the performance of the ranking models across the four binary metrics defined in~\cref{sec:objective-defs} (i.e. if a track is completed, if it is a discovery, etc.), reporting both NDCG@5 and NDCG@10.
mAP is also used for some analysis of the behaviour of these systems.

\vspace{-1mm}

\subsection{Comparison across Methods}

\cref{tab:results} lists the performance across our objectives for all the models.
Focusing on single objective methods optimised for user satisfaction, we see that set-level models perform on par with track-level models, while encoding the full input set at once (rather than each input sequentially before ranking them). 
Moreover, as expected, learnt models far outperform the non-learnt relevance ranker. 

Looking at set-level methods, we see that both $\setrank$ and SetTRM2GRU achieve competitive performance. 
However, an autoregressive model, such as a GRU, is prohibitively difficult to deploy in large-scale scenarios where fast processing is a crucial requirement. 
Hence, due to the practical limitations of DNNs and autoregressive models, we set $\setrank$ as our best user-satisfaction baseline.

Focusing on multi-objective methods, we observe that naively training learning-to-rank methods on composite multi-objectives significantly drops performance in \sat (-15\% in \ndcgfive).
In addition, designing the right composite label to train multi-objective learning-to-rank methods is non-trivial, especially without prior information on how different objectives co-interact.

Our inference-based baseline \mostrasum, which performs a simple weighted sum of the beams, observes large fluctuations given the weight $\alpha$.
Compared to $\setrank$, it either achieves gains in user satisfaction, with drops on artist- and platform-centric objectives, or gives up to 20\% gains in \ismid and \discovery metrics but severely hurts user satisfaction (-10\% in the \sat metric).

Finally, we observe that our \mostra method with counterfactual and submodular scoring affords much finer control than \mostrasum over model performance across all four metrics. 
Indeed, the counterfactual scoring in \mostra lets us control the trade-off across objectives, and gives significant gains across \disco, \promo and \ismid objectives.
\mostra achieves these gains while maintaining performance on \sat that is on par with the best single objective ranker, and slightly higher than $\setrank$.


\subsection{Performance by Competition within Sets}

\begin{figure}
    \centering
    \vspace{-3mm}
    \includegraphics[width=0.85\linewidth, trim={0cm 0cm 0cm 0cm}, clip]{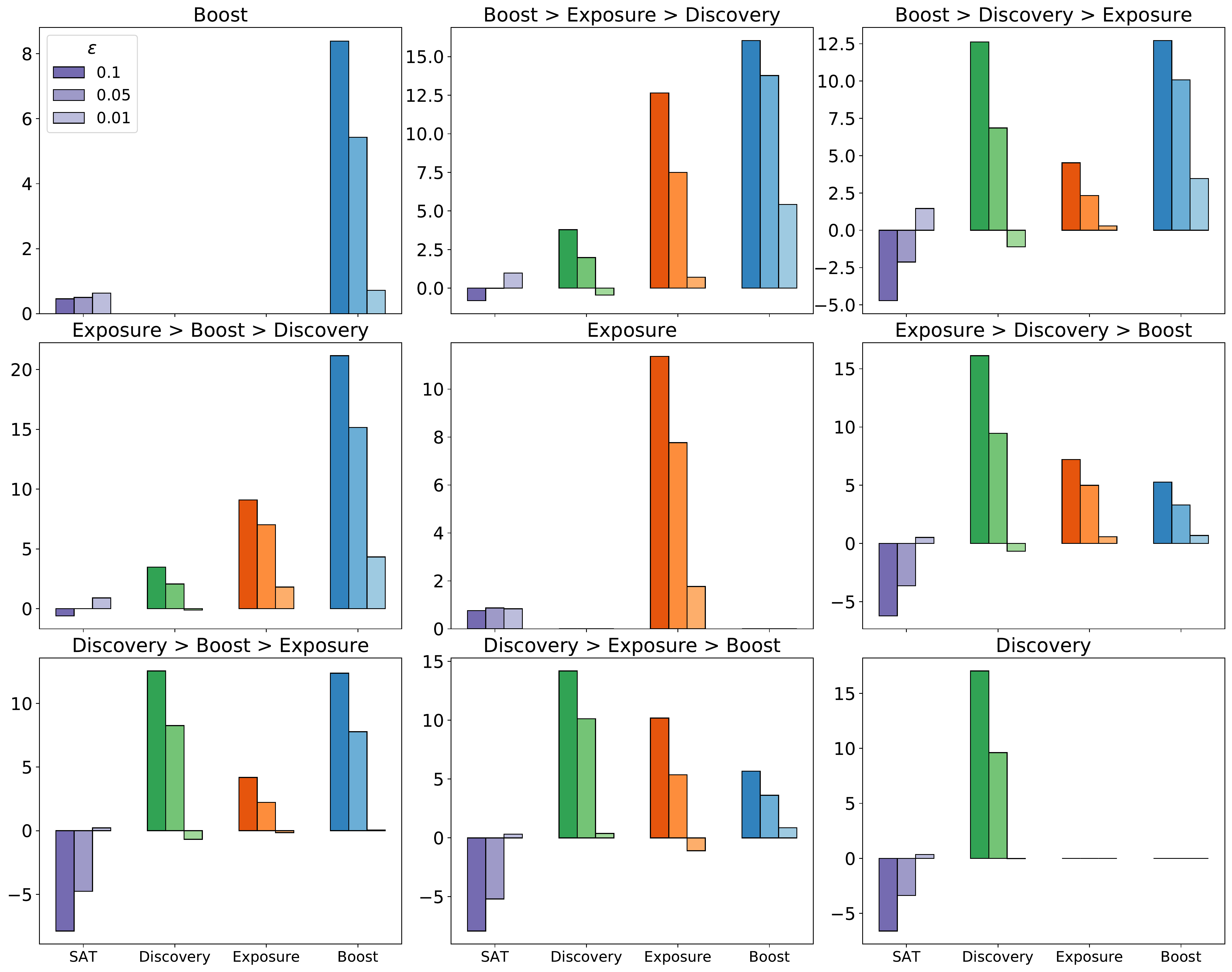}
    \vspace{-3mm}
    \caption{\small{Performance gains by competition in session. Top-left: gains in sessions with only \promo non-user objective. Top-centre: gains in sessions with more tracks for \promo, than for \ismid, than for \disco; etc.}}
    \label{fig:competition}
    \vspace{-6mm}
\end{figure}

Earlier in \cref{fig:heatmap}(d), we demonstrated the vast diversity of session types in our data: some sessions only have tracks belonging to one of these objectives, while various sessions have tracks belonging to all objectives. 
\cref{fig:competition} presents a $3\times3$ plot of the gains of \mostra w.r.t. $\setrank$ in terms of \ndcgfive for different types of sessions.

In the diagonal, we find sets for which only one type of non-user objective tracks are available. 
We observe that for different values of the counterfactual degree $\epsilon$, \sat only drops in sets with \disco. 
This is in line with the results in~\cref{fig:heatmap}, where we saw a negative correlation between \sat and \disco. 
For both \promo-only and \ismid-only sets, we observe significant gains in the respective metrics, while also seeing \sat metric increasing. 
Similarly, whenever \disco are either the most or second-most frequent type of non-user objective in a session, we observe drops in \sat scores. Moreover, \promo scores tend to receive higher gains across the different types of sets, a result consistent with the boxplot in \cref{fig:heatmap}(e) where \promo had shown higher historic average \sat.

\subsection{Balancing Trade-offs: Relaxing Constraints}

The flexible scoring of \mostra allows us to control the trade-off between objectives in different scenarios. For instance, we consider three different scenarios: (i) prioritise short-term \sat, (ii) additionally focus on \ismid of emerging artists, or (iii) consider on all the objectives defined above. 
\mostra  allows us to tackle each scenario on-the-fly thanks to its flexible decoding component.

Focusing on short-term satisfaction, compared to SetRank, \mostra gains more than 1 NDCG@5 point in \sat by surfacing fewer \disco tracks (-1.4 points), with no significant changes over the two other metrics. When asked to additionally optimise \ismid tracks, \mostra ($\epsilon = 0.05$) gains 6 NDCG@5 points whilst also increasing \sat (1 point); performance on \promo is unaffected while keeping -1.4 points in \disco. Finally, in the challenging case where all the objectives are optimised, we observe gains in each of them, but with a minimal drop in \sat (as shown in \cref{tab:results}).

\begin{figure}
    \centering
    \vspace{-3mm}
    \includegraphics[width=0.8\linewidth, trim={0cm 0cm 0cm 0cm}, clip]{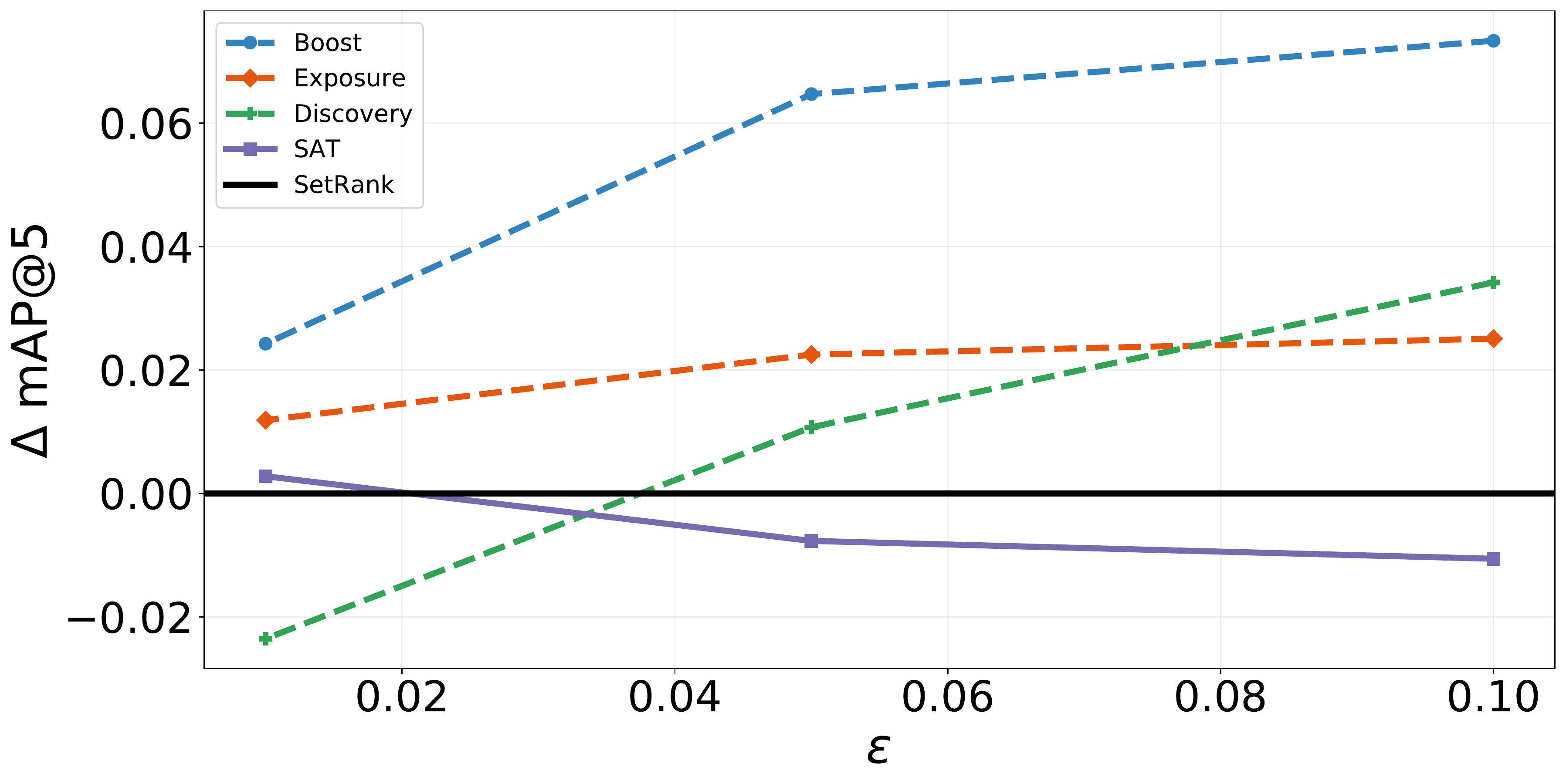}
    \vspace{-5mm}
    \caption{Performance boost across relaxation thresholds. $\epsilon$ controls the allowed drop in \sat.} 
    \label{fig:diversity}
    \vspace{-7mm}
\end{figure}

In \cref{fig:diversity}, we illustrate the benefits offered by \mostra's flexible design: we increase the $\epsilon$ relaxation threshold, allowing the model to consider a wider set of tracks, potentially satisfying other objectives beyond user satisfaction. For each objective, we plot the mAP@5, that shows how many tracks from different objectives we are able to surface in the top-5 recommendations. We only consider tracks that are streamed, and hence successful for user satisfaction objective.

\mostra ($\epsilon=0.01$) results in more \promo and \ismid streamed tracks than SetRank. This also leads to a higher \sat score but at the expenses of streaming fewer \disco tracks.
As we increase $\boldsymbol{\epsilon}$, then, the model is able to increase other objectives, albeit at the cost of satisfaction, but in a more controlled way than the \mostrasum baseline. 
Also, it is interesting to observe that when a large-enough subset of multi-objective tracks is available at each step ($\epsilon=0.1$), the ratio of \disco is higher than that of \ismid. Given that sessions have more streamed \ismid tracks than \disco (\cref{fig:all_good}) tracks, this shows that our submodular function (\cref{eq:submod}) can balance across objectives and provide diverse recommendation of tracks from multiple objectives.

\subsection{Measuring Success across Objectives}

Compared to $\setrank$,
 \mostra gains ~1.2\% \ndcgfive on user satisfaction whilst also recommending tracks that satisfy all other objectives, gaining 10\%, 3\% and 8.5\% on \disco, \ismid and \promo metrics, respectively. 

We wish our model to not only show artist- and platform-centric tracks, but also show tracks users would indeed stream. 
To this end, we differentiate between cases where the user listens to a track that meets non-user-centric objectives (denoted as \textit{Good}) to cases where such a track is just shown to user (regardless of whether they stream it or skips it) (denoted as \textit{All}). If we consider the \discovery objective, we refer to all discoveries surfaced to users as \textit{All} and the discoveries actually streamed by the user as \textit{Good} discovery tracks. A higher \textit{Good}--\textit{All} ratio indicates that the model is able to show better discovery candidates to users. 

\begin{figure}[t]
    \centering
    \vspace{-4mm}
    \includegraphics[width=0.9\linewidth, trim={0 0 0 0}, clip]{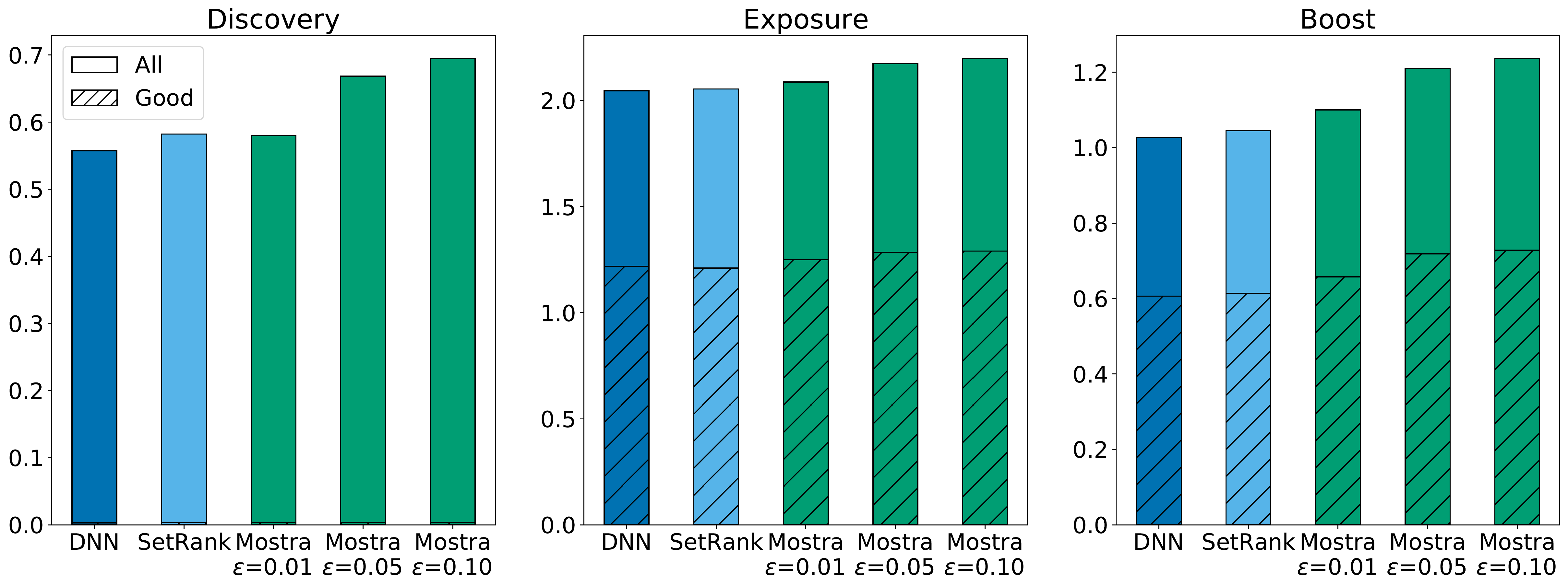}
    \vspace{-4mm}
    \caption{\emph{Good}--\emph{All} platform-centric tracks performance.} \label{fig:all_good}
    \vspace{-4mm}
\end{figure}

\cref{fig:all_good} shows the average number of top-5 \disco, \ismid and \promo tracks shown to the users by different methods, with the shaded regions denoting the proportion of these that were actually streamed (i.e. \textit{Good}--\textit{All} ratio).
We see that \mostra is able to surface more tracks for each of these objectives, and is doing so without hurting user satisfaction metric (\cref{tab:results}). 
Also, for the other objectives, \mostra can recommend more tracks and of higher quality, since we observe an increase not only in the amount of \emph{All} and \emph{Good} tracks shown, but also in their ratio.

\subsection{Performance Generalisation}

We further investigate the performance of \mostra across session lengths, and session difficulty, and present broad trends observed. We find evidence that \mostra ($\epsilon=0.05$) surfaces more \ismid and \promo tracks to users in longer sessions (+10\% and +15\% \ndcgfive, respectively), keeping \disco constant.
That is, \mostra can exploit larger sets to meet multiple, non-competing objectives with no effect on user satisfaction.

To understand performance by session difficulty, we compute the average ground-truth \sat score and use this to group sessions based on their difficulty. The higher the skip rate in a session, the more \emph{difficult} the session.
We observed that \mostra ($\epsilon=0.01$) performs better than $\setrank$ across all difficulty level, and as the sessions become easier, the gains offered by \mostra model also increase and can be controlled with~$\epsilon$. 

Finally, to understand generalisation across encoding approaches, we applying the proposed beam search decoding to \fc$_{\mathrm{RMSE}}$, the model that achieved higher satisfaction score, and find similar performance trends of gains in other objectives with minimal loss in satisfaction.
This shows that the proposed multi-objective decoding module is model-agnostic and can boost multi-objective performance across different deployed systems.

\vspace{-1mm}

\section{Conclusion} \label{sec:conclusion}

We focused on the problem of multi-objective music recommendations, typical of multi-stakeholder platforms.
Our analysis revealed the complex interplay that exists between different objectives, showing the potential and difficulty of jointly satisfying them.

Motivated by the practical real-time requirements of large-scale multi-stakeholder platforms, we then formulated the problem of just-in-time, multi-objective optimisation.
We proposed \mostra, an end-to-end neural architecture that can easily be adapted to meet the dynamic strategic needs through multi-objective recommendations.
Our framework affords system designers the ability to control, on-the-fly, the trade-offs across objectives when targeting multiple, competing objectives. 
This flexibility is given by a novel beam search algorithm, which can balance multiple objectives in a controlled manner via counterfactual decisions.
Experiments on a large-scale dataset of music streaming showed that our method achieves competitive performance on user satisfaction compared to state-of-the-art methods, while also significantly increasing gains on the other artist- and platform-centric objectives.

We believe that our findings and our problem formulation have implications on the design of recommender systems for multi-stakeholder platforms. 
We envision future extensions to include multiple prediction objectives, including global objectives that are not session-specific but aggregated over time, and encoder-decoder architectures that are jointly trained on such objectives to better address evolving business needs.

\vspace{-2mm}

\clearpage
\balance

\bibliographystyle{ACM-Reference-Format}
\bibliography{references}

\clearpage
\appendix
\section{Experimental Details} \label{sec:exp-details}

For each user's radio session, we log all tracks the user interacted with, along with user interaction signal for each track (e.g. whether it is skipped). 
\cref{tab:featurestable} lists the logged features for interacted tracks.

\begin{table}[htp]
    \centering
    \footnotesize
    \caption {Description of user, track, and user-track combination features used in the neural rankers.} \label{tab:featurestable} 
    \vspace{-3mm}
    \resizebox{\columnwidth}{!}{%
    \begin{tabular}{l|c|l}
    \toprule
    \textbf{Type}                & \textbf{Feature} & \textbf{Description}                                        \\ \midrule
    \multirow{2}{*}{User}        & embedding        & 40 dimensional learnt word2vec vector of user               \\ \cline{2-3} 
                                 & country          & country of registration for user                            \\ \hline
    \multirow{5}{*}{Track}       & embedding        & embedding \& 40 dimensional learnt word2vec vector of track \\ \cline{2-3} 
                                 & popularity       & normalised popularity of the track                          \\ \cline{2-3} 
                                 & genres           & genres relevant to the track                                \\ \cline{2-3} 
                                 & acoustic         & 16 derived acoustic features                                \\ \cline{2-3} 
                                 & track length     & track duration in seconds                                   \\ \hline
    \multirow{3}{*}{User--Track} & similarity       & user--track embeddings cosine similarity                    \\ \cline{2-3} 
                                 & distance         & user--track embeddings Euclidean distance                   \\ \cline{2-3} 
                                 & genre affinity   & affinity for highest user--track overlapping genre          \\ \hline
    Session                      & session ID       & unique session identifier for learning embeddings           \\ 
    \bottomrule
    \end{tabular}%
    }
\end{table}

We compare \mostra with three groups of state-of-the-art recommendation models. 
The first group corresponds to track-level methods, which focus on predicting scores one track at a time. 
These single-objective methods solely focus on user satisfaction, but which might surface tracks from other objectives organically, without explicit optimisation. 
\begin{description}
    \item [Relevance ranker:] Ranks tracks by estimating user--track relevance scores one track at a time, and sorts the estimated relevance scores. We learn contextual embeddings for each user and track, and compute their cosine similarity.

    \item [\fc:] A track-level model to predict user--track interaction, trained using a 5-layer fully-connected architecture with ReLU activations in between.
    The layer widths are [256, 1024, 2048, 1024, 256] and were chosen so as to have similar size ($4.8$M) as our model ($5.6$M). The network was trained with either RMSE, BCE or \attnrank{} loss functions. These  models constitute well-established neural baselines~\cite{45530,10.1145/3038912.3052569,10.1145/2988450.2988454}.
\end{description}
The second group are single-objective set-level models and model interaction effects across all tracks in the set and output a permutation. 
These models were trained using the \attnrank{} objective.
\begin{description}
    \item [SetTRM2GRU:] A $\trm$ encoder with a GRU decoder.
    The encoder consists of 6 layers, 8 attention heads, 20 induced points, hidden size of 256, and intermediate size of 1024.
    The decoder has 1 layer with hidden size of 256, decoder size of 512, and additive attention of size 128. Size: $7.3$M parameters.
    \item [SetTRM2TRM:] A $\trm$ encoder with a TRM decoder.
    The decoder here has 1 layer, 8 attention heads, 20 induced points, hidden size of 256, and intermediate size of 1024. Size: $6.8$M.
    \item [SetRank:] A state-of-the-art recommender system based on $\trm$ that sorts a set by predicted relevance scores~\cite{10.1145/3397271.3401104}.
    In our framework, it can be viewed as applying a \emph{sort} function to the input tracks according to the scores given by \cref{eq:enc}.
    The $\trm$ is the same as above. Model size is $5.6$M parameters.
\end{description}
Finally, the third group is made of multi-objective sequencing methods (including variants of the proposed method), which operate at the set-level and consider multiple objectives.
\begin{description}
    \item [\moltr:] A learning to rank model, which extends $\setrank$ to multi-objective training. Multi-objective labels (i.e. \promo, \ismid and \disco) are given as input to the model, which is then trained to predict the sum of \sat and the other objectives when the track is completed: $s_i = \sat_i + \sat_i\sum_{j=1}^J \mathbf{E}_{ij}$.
    This is our learnt multi-objective baseline. 
    \item [\mostrasum:] A first, non-counterfactual version of our \mostra method, in which multi-objective labels are given at inference. The model ranks tracks according to the weighted sum of all the objectives whenever a track has a predicted relevance score higher than $0.5$, hence mimicking \moltr{} but as an inference method: $s_i = \scores_i + \mathds{1}\{\scores_i > 0.5\} \alpha\sum_{j=1}^J \mathbf{E}_{ij}$.
    This is our online multi-objective baseline. 
    \item [\mostra:] Our $\trm$-based model with multi-objective counterfactual decoding (\cref{sec:model}).
\end{description}
Every model was trained using the Adam optimiser with a learning rate of 0.0005 and decay factor of 0.1.
We batch 128 sequences padded to a maximum length of 100 tracks and train each model for 12,500 steps on a single Tesla T4 GPU.
Model selection was based on the best validation performance.
We use Normalised Discounted Cumulative Gain (NDCG)~\cite{10.1145/582415.582418} to evaluate the performance of the ranking models across the four binary metrics defined in~\cref{sec:objectives} (i.e. if a track is completed, if it is a discovery, etc.), reporting both NDCG values at the ranks of $5$ and $10$.
Mean average precision (mAP) is also used for some analysis of the behaviour of these systems.

\end{document}